\documentstyle[preprint,aps]{revtex}
\preprint {IMSc/99/01/02}
\begin{document}
\draft
\title{Exclusion statistics: A resolution of the problem of negative 
weights}

\author{ M. V. N. Murthy and R. Shankar}

\address
{The Institute of Mathematical Sciences, Chennai 600 113, India.\\
}
\date{\today}
\maketitle
\begin{abstract}

We give a formulation of the single particle occupation probabilities for
a system of identical particles obeying fractional exclusion statistics of
Haldane. We first derive a set of constraints using an exactly solvable
model which describes an ideal exclusion statistics system and deduce the
general counting rules for occupancy of states obeyed by these particles.
We show that the problem of negative probabilities may be avoided with
these new counting rules. 
  
\end{abstract}


\narrowtext

\section{Introduction}

A few years ago Haldane\cite{hald} proposed a generalized exclusion
principle in which adding particles in to a system leads to a change in
the dimension of the single particle space. Specifically, the generalized
exclusion principle envisages systems in which the addition of one
particle blocks off $g$ single particle states for the others, where $g$,
the fractional exclusion statistics(FES) parameter is an arbitary number. 
Obviously $g=0$ for bosons and $g=1$ for fermions. This leads to the
following formula for $D_N(g,d)$, the dimension of the N particle Hilbert
space, if the dimension of the single particle space is d,
\begin{equation} 
D_N(g,d)~=~{(d+(1-g)(N-1)! \over N!(d-1-g(N-1)!}
\label{haldane} 
\end{equation} 
which reduces to the familiar expressions for Bose and Fermi statistics
for $g=0$ and $g=1$. 

Thermodynamic properties of an ideal gas of exclusion particles have been
investigated widely. Specifically, a definition of an ideal
gas of particles with nontrivial exclusion statistics was given in
references \cite{wu,isa}. In this definition it was assumed that if there
were $d$ levels of energy $\epsilon$, then the dimension of the Hilbert
space with N particles of energy $\epsilon$ is given by equation
(\ref{haldane}). The distribution function may then be
computed\cite{wu,isa,raj,ouvry} and is given by
\begin{equation}
n(\epsilon)~=~{1 \over (w(\epsilon)+g)},
\end{equation}
where $w(\epsilon)$ is the
solution of the equation\cite{wu,ouvry}
\begin{equation}  
w(\epsilon)^g(1+w(\epsilon))^{(1-g)}~=~e^{\beta\epsilon} 
\label{wu} 
\end{equation}
and $\beta$ is as usual the inverse temperature. If we attempt to
interpret this distribution function as arising from the statistical
mechanics of a single mode with the statistical weight $\rho_n
e^{-n\beta\epsilon}$ for the mode to be occupied by $n$ particles, then it
was found \cite{wilnay,poly} that some of the $\rho_n$s are invariably
negative if $g$ is different from 0 and 1. This raises the problem of
interpreting these negative probabilities. It has also been speculated
that these negative probabilities are an essential feature of nontrival
exclusion statistics \cite{poly}. 

A crucial property of exclusion statistical interactions is that they
should cause shifts in single particle energies at all scales\cite{ms1}.
This property is realized by a large class of one dimensional models of
interacting fermions where Fermi liquid theory breaks down \cite{pwa,ms2}.
In fact it has been shown exactly that quasiparticles with nontrival
exclusion statistics exist in a class of models that are solved by the
Bethe ansatz \cite{isa,bwu}. In particular the quasiparticles of the
Calogero-Sutherland model (CSM) behave like ideal exclusion statistics
system \cite{isa,bwu,ms3,ha}. A feature of the exclusion statistics as
gleaned from the analysis of various models is that the exclusion acts
across a set of levels unlike in the case of Fermi or Bose statistics
where the exclusion principle is stated with a single level in mind. It is
this crucial difference that results in the occurance of negative
probabilities.  We will show that the particles obeying fractional
exclusion statistics can be characterised by constraints on the sets of
occupation numbers. There are no negative probabilities if these
constraints are obeyed. If these constraints are relaxed then the negative
weights arise in order to compensate for the resulting over counting.
Indeed this is the way we encounter negative probabilities in other
systems in physics- for example in gauge theories, they arise in the ghost
sectors. Ghosts come from the Jacobian associated with nonlinear gauges
which essentially ensure the correct counting of states.  Another example
is that of the Wigner distribution function in quantum mechanics which is
not positive definite precisely because some constraints are relaxed.  A
formulation based on the variable number of single particle states, which
depends on the total number of particles in the system, has been discussed
by Isakov\cite{isakov} as a way to avoid the problem of negative weights.
Recently, a microscopic interpretation of exclusion statistics systems has
been advanced by Chaturvedi and Srinivasan\cite{subash} where they show
how this problem of negative probabilities may be solved for semions,
$g=1/2$. They have also indicated how their method may be generalised to
other values of $g$. 

In this paper we first discuss the origin of negative probabilities in
exclusion statistics particle systems. To do this we have chosen an
unusual starting point in an equation and its solution given by
Ramanujan\cite{ram}. This starting point makes precise the statements
about the occurence of negative probabilities.  We then formulate a
counting principle based on the set of constraints which reproduces the
Haldane dimension formula. We first extract the counting rules starting
from an exactly solvable model of interacting particles and state them in
the form of counting rules for arbitrary systems obeying exclusion
statistics. This method not only avoids the negative probabilities, but
with minimal modification reproduces the results derived by Chaturvedi and
Srinivasan for the semion. The counting principle is however not
restricted to semions alone. 

\section{The problem of negative weights}

The problem of negative probabilities was first pointed out by Nayak and
Wilczek\cite{wilnay} and elaborated by Polychronokos in a recent
paper\cite{poly}. In order to clarify the origin of negative probabilities
or weights, we first discuss an equation and its solution due to Ramanujan
\cite{ram}.  Ramanujan considered the following equation: 
\begin{equation} 
aqX^p-X^q+1=0, 
\label{ram1} 
\end{equation} 
where $a$ may be complex and $p,q$ are positive.The general 
solution for $X^d$ is, 
\begin{equation} 
X^d~=~ \sum_{N=0}^{\infty}C_N(p,q,d)a^N, 
\label{ram2} 
\end{equation} 
where $C_0(p,q,d)=1$ and $C_1(p,q,d)=d$ and

$$C_N(p,q,d)={d \over N!} \prod_{j=1}^{N-1}(d+Np-jq),~~~N\ge 2.$$ 

To make connection with the result obtained by Polychronokos\cite{poly},
which is a particular case of the general solution given by Ramanujan, we
now put $p=(1-g)$ and $q=1$, then 
\begin{equation} C_N(1-g,1,d)~=~d
{(d+(1-g)N-1)! \over N!(d-gN)!} 
\end{equation} 
which is clearly different from the dimension formula of Haldane. 
However, it correctly reproduces the bosonic and fermionic dimension
formula for $g=0$ and $g=1$ respectively. This dimension formula was
derived independently by Polychronokos\cite{poly} with the restriction
that any two particles are atleast $g$ sites apart when placed on a
periodic lattice. One can also derive the Haldane dimension formula with
the restriction that any two particles are $g$ sites apart but without the
restriction of periodicity. 

Further if we put $X=(1+w^{-1})$ and $a=e^{-\beta \epsilon}$ in
eq.(\ref{ram1}), we immediately obtain equation (\ref{wu}) derived
earlier by Wu. The important point to notice here is that the dimension
formula that precisely leads to the distribution function derived earlier
\cite{wu,isa,raj,ouvry} is given by $C_N$ and not the Haldane dimension
formula.  In the limit $d>>1$, however, it is easy to see that
$$C_N(1-g,1,d)=D_N(g,d) + O({1 \over d}).$$ 
Therefore in the continuum limit, the $C_N$ and $D_N$ are approximately
the same. 

The grand canonical partition function of the system may be 
written as,
\begin{equation}
Z~=~(1+w^{-1})^d~=~\sum_{N=0}^{\infty} C_N(1-g,1,d)e^{-\beta N\epsilon},
\label{partfac}
\end{equation}
where $w$ satisfies equation (\ref{wu}). We have also assumed that all
the energy levels are degenerate with energy given by $\epsilon$. Note
that this is an exact expression and no assumption is required on the
single particle dimension $d$. The negative weights arise
\cite{wilnay,poly}, when one insists on expanding $1+w^{-1}$ in powers of
$e^{-\beta\epsilon}$. From equation (\ref{ram2}) and the definitions
following the equation, it follows that,
\begin{equation}
1+w^{-1}~=~\sum_{n=0}^{\infty} C_n(1-g,1,1)e^{- \beta n\epsilon}
\label{part1mode}
\end{equation}
The weights 
\begin{equation}
C_n(1-g,1,1) = ~p_n~=~\prod_{m=2}^{n}(1-\frac{gn}{m})
\label{weights}
\end{equation}
are always negative for $gn>m$ for some $m$\cite{poly}. This is indeed the 
problem of
negative weights associated with exclusion statistics and is claimed to be
inherent in the exclusion statistics.  There are however a few points to
note: The negative probabilities arise because of
our insistence on the factorization\cite{subash} implied in
eq.(\ref{partfac}).  For example, combining equations (\ref{partfac}) and
(\ref{part1mode}) we have,
\begin{equation}
Z~=~\sum_{\{n_j\}}(\prod_j C_{n_j}(1-g,1,1))e^{-\beta\epsilon \sum_j n_j},
\label{partuncons}
\end{equation}
where the sum is an unconstrained one over all sets of occupation numbers. 
The over counting resulting from this unconstrained sum is compensated by
the occurence of negative weights. We next derive the precise counting 
rules which impose constraints on this sum and avoids this problem. 

\section{Realization in an interacting system and counting rules}

Any  realization of fractional exclusion statistics must have its
origins in systems of interacting particles. The expectation is that under
certain conditions systems of interacting particles which obey Fermi or Bose
statistics may be described in terms of quasiparticles (or quasiholes)
which obey fractional statistics. The quasiparticles of the CSM 
behave like ideal exclusion statistics particles. The main feature of CSM
is that the total energy of the many-body system can be written in terms
of single quasi-particle energies which involve shifted momenta and these
shifts contain the information about the exclusion statistics of the
quasiparticles. In this section we analyse these shifted momenta and make
explicit connection with the formula in eq. (\ref{haldane}). We then
use them to obtain constraints on the allowed set of occupation numbers.
These are what we refer to as the counting rules that reproduce the
formula in eq.(\ref{haldane}). The statistical mechanics of the
system obeying these constraints is then the same as that defined by Wu
\cite{wu} and all statistical weights are positive. 
    
We begin with the trignometric Sutherland 
model\cite{suth1} of an  
N-particle  system on a ring of unit radius. The Hamiltonian is given by, 
\begin{equation}
H = -\sum_{i=1}^N \frac{\partial^2}{\partial x_i^2} + 
\sum_{j<i}\frac{2g(g-1)}{\sin^2[(x_i-x_j)/2]},
\label{csmham}
\end{equation}
where $g$ is the interaction parameter. We will soon identify this with 
the statistical parameter of the exclusion statistics. While the model 
can be applied to both interacting bosons and fermions, we choose to work 
in the fermionic basis here after. The energy 
of an N-fermion state may be written in terms of shifted momenta as
\begin{equation}
E = \sum_{i=1}^{\infty} k_i^2 n_i,
\end{equation}
where $n_i=0,1$ and the shifted momenta $k_i$ (also called pseudo 
momenta in Ref.\cite{ha}) are given by
\begin{equation}
k_i~=~m_i - (1-g){(N_i^- - N_i^+) \over 2},
\label{shifmom}
\end{equation}
where $m_i$ are distinct integers, 
$N_i^{-(+)}$ are the number of particles with shifted momenta 
less (greater) than $k_i$. Note that we could have also started with the 
Calogero-Sutherland model with harmonic confinement. The results below 
follow analogously with the proviso that we have shifted energies instead
of shifted momenta. 

First we establish the relationship between the shifted momenta given
above and the Haldane's dimension formula (\ref{haldane}). Consider the
above system with an upper and lower cutoff on the momenta, $k_{max}$ and
$k_{min}$ respectively. We divide this range of momenta into cells of unit
length (the first and last cells could be smaller) and define the
occupancy of the $j^{th}$ cell, $n_j$ to be the number of particles with
momenta $k_i$, such that $j+1>k_i\geq j$. We identify single particle
space dimension $d$ with the number of cells in the range, i.e
$d=k_{max}-k_{min}$, where $d$ may be fractional. If we now denote
the range of the $m_i$s by $d_F$, we have
\begin{equation}
d_F = m_{max}-m_{min} = d+(1-g)(N-1).
\end{equation}
Since there exists an $m_i$ for every $k_i$, the total number of states
in the range $k_{max}~-~k_{min}$ is the same as that between 
$m_{max}~-m_{min}$.
The total number of states is then the number of ways N distinct 
integers can be picked from $d_F$ distinct integers, i.e $^{d_F}C_N$, 
as in fermionic description. Substituting for $d_F$ from the above
expression we immediately reproduce the Haldane dimension formula in  
eq.(\ref{haldane})\cite{ms1}. 
 
In order to obtain the counting rules we will first derive three
properties of the set of momenta $\{ k_i \} $.  If $k_i$ are ordered such
that they increase with increasing $i$, then we have,
$k_{i+1}-k_i=m_{i+1}-m_i-(1-g)$.  If $g<1$, then it follows that
$m_{i+1}>m_i$.  Further, if $m_{i+1}-m_i=1$ then $k_{i+1}-k_i=g$ and if
$m_{i+1}-m_{i}>1$ then $k_{i+1}-k_{i}>(1+g)>1$ because $m_i$'s are
integers. 

We can then draw the following three conclusions from the 
properties of the shifted momenta $k_i$:
\begin{enumerate}
\item 
The ordering in $k_i$s is the same as   the
ordering in $m_i$s. 
\item "Close packed" $m_i$s with unit spacing correspond to 
"close packed" $k_i$s with spacing $g$. 
\item The gaps between any 
two non-close packed $k_i$s is greater than 1.
Therefore all the $k_i$s in any cell are close packed. 
\end{enumerate}

We now come to the question as to what are the constraints on the sets of
occupation numbers $\{k_i\}$. For example, if $g=0$, there are no
constraints as in the bosonic case. If $g=1$ the constraints are $n_j 
\leq 1$ as in
the case of fermions. For any other $g$, one obvious constraint come from
the second property derived above, namely the occupancy of the j-th cell 
$n_j \leq {1 \over g}$ which
specifies the maximum occupancy of a given cell assumed to be of unit
spacing. This is the same constraint one derives from the distribution
function of Wu (\ref{wu}). An important departure from the usual bosonic
and fermionic case is that the cell size is important and cannot be
arbitrarily taken to zero as in the case of bosons and fermions
\cite{wilnay}. 

There are further constraints on the occupancy. To formulate them we use 
the third property. Let $k_L$ be the lowest momentum in the $j^{th}$ 
cell. Then from the second and third property, it follows that
\begin{equation}
k_L + g(n_j-1) < j+1 .
\label{kL}
\end{equation}
We can write $k_L$ as $k_L=j+f(k_L)$, where $f(k_L)$ denotes the 
fractional part of $k_L$, that is,  $0 \leq f(k_L) < 1$. We then have,
\begin{equation}
f(k_L)+g(n_j-1)~<~1 .
\label{cons1}
\end{equation}
From equation (\ref{shifmom}), we can express $f(k_L)$ as a function 
of the occupation numbers,
\begin{equation}
f(k_L)~=~f\left [-(1-g){[N^-_{cj}-N^+_{cj}-(n_j-1)] \over 2}\right ],
\label{cons2}
\end{equation}
where $N^-_{cj}=\sum_{l<j} n_l$ and $N^+_{cj}=\sum_{l>j} n_l$. Equations 
(\ref{cons1}) and (\ref{cons2}) then constitute a set of constraints on 
the occupation numbers.

We will now show that these form a complete set of constraints.  Namely,
given any set of occupation numbers, $\{n_j\}$, that satisfies the
constraints, there exists a set of momenta, $\{k_i\}$, that realizes it.
To do this, consider a set $\{n_j\}$, where $j_{min} \leq j \leq j_{max}$.
The lowest value of the momentum in the $j^{th}$ cell is uniquely
determined by the occupation numbers through equation(\ref{cons2}).
Because of the third property, all the other momenta are also uniquely
determined.  Hence we have shown that there are no more constraints. 
Equations (\ref{cons1}) and (\ref{cons2}) form a complete set of
constraints. Note also that the above logic implies that there is a one to
one correspondence between the sets of occupation numbers, $\{n_j\}$, that
satisfy the constraints (\ref{cons1}) and (\ref{cons2})and the sets of
momenta, $\{k_i\}$, that satisfy equation (\ref{shifmom}). 

We can now remove the scaffolding of the Sutherland Model that we 
started with and $\it 
define$ exclusion statistics system by the above constraints. The 
connection to the dimension formula in eq.(\ref{haldane}) established 
earlier implies that
\begin{equation}
\sum_{\{n_i\}}F(\{n_i\})~=~D_N(g,d), 
\label{consdim}
\end{equation}
where $N=\sum_j n_j$ and $F(\{n_i\})=1$ if $\{n_i\}$ satisfy the 
constraints and zero otherwise. Note that the weights now are positive 
definite. There are no negative weights once the constraints are imposed. 

Next, we construct some simple examples from the above counting rules. 
For simplicity we look at occupation numbers for special values of $g =
1/m$ where $m$ is an integer.  The rules formulated above for the
occupation number of exclusion particles may be combined and restated
thus: 

\begin{quote} 
{\it  Let $m = 1/g$, and let
$N_i$ be the number of particles in the occupied states below some
$i$th level, $N_i=\sum_{j<i}n_i$. Then an occupation $n_i (n_i \le m)$ is 
allowed iff $(N_i~ mod~ m) \le (m-n_i)$ }. 
\end{quote}
This rule now includes all the three constraints stated above.  

To see how this rule is implemented, consider a system of N-particles 
spread over $d$ states. In
order that $D_N$ is an integer, we choose $N=mp+1$, where $p$ is an
integer. Since $ N \le md$, we have $p<d$.  We shall divide these $d$
states into cells. An allowed configuration may be represented as a 
string of numbers $(n_1n_2n_3,...)$, where each $n_i \le m$ denotes the 
occupancy of levels ordered from left to right. Instead of dealing with a 
configuration where all $N$ particles are spread over $d$ states (some 
which may be empty), we can simplify the discussion by considering one 
cell at a time. Each cell may now have a partition of $m$. This allows us 
to fill the subsequent cells without reference to the previous cell 
according to the counting rules since $N_i ~mod~ m =0$. We now fill 
each cell with a partition of $m$ which is allowed by the rules given 
above. This then generates all possible allowed configurations whose sum 
is given by $D_N$. 

If, in particular, we are interested in expectation values of symmetric 
functions of $n_i$, we can work with symmetrised weights. Consider a 
symmetric operator $O(\{n_i\})$. The expectation value of this operator 
may be written as,
\begin{equation}
<O(\{n_i\})> = 
\frac{\sum_{\{n_i\}}F_s(\{n_i\})O(\{n_i\})}{\sum_{\{n_i\}}F_s(\{n_i\})}, 
\label{expect}
\end{equation}
where
\begin{equation}
F_s(\{n_i\})~=~\frac{1}{d!}\sum_{p}F(p_{n_i}). 
\end{equation} 
Here $p$ stands for all permutations of the allowed configurations. Every 
allowed configuration in $\{n_i\}$ may be characterised by  
the multiplicities $q_n$, namely a given allowed configuration may be 
written as a string, $m^{q_m} (m-1)^{q_{m-1}}...1^{q_1}$, where 
$q_1 +2q_2 +...+mq_m = N$. We may now also allow any permutation of these 
occupancies (with zeros added to make up d-states). The dimension of the 
N-particle space may then be written as 
\begin{equation}
D_N(g,d)= \sum_{\{q_n\}} f^N_m(q_1,q_2,...)~~~^dC_q,
\end{equation}
where $q=\sum_{n=1}^{m}q_n$. The new weights $f$ are defined as, 
\begin{equation} 
f^N_m(q_1,q_2,...q_m) = \frac{M_a(q_1,q_2,...q_m)}{M_t(q_1,q_2,...,q_m)},
\label{fweight}
\end{equation}
where
$M_a$ allowed configurations after symmetrising and $M_t$ is the total 
number of configurations for a given set of $q$'s which define a 
configuration. We shall clarify this now with specific examples.

\subsection{The case of semion ($g=1/2,m=2$)}

The maximal occupancy of a state in this case is 2. Hence allowed
occupancy of a state is 2 or 1. Zeros may occur any where without changing
the rules. Let us implement this in the specific case of $d=4, N=5$, say.
In this case the allowed configurations are given by the strings
(2210),(2111),(1121). In the first configuration, zero can be anywhere and
therefore there are four configurations.  Notice that a string of the form
(1211) or (1112) violates the counting rules. Therefore counting all the
allowed configurations we obtain $D_5(1/2,4)=6$. This is exactly what one
gets from the Haldane formula. 

Further if we symmetrise each of these allowed configurations, then the new 
weights may be computed using eq.(\ref{fweight}). 
In the specific case of $m=2$, we have
\begin{equation}
M_t(q_1,q_2) = ^{q_1+q_2}C_{q_2}, ~~~~ 
M_a(q_1,q_2) = ^{p}C_{q_2},
\end{equation}
where $p$ is defined through the equation $N=2p+1$.  The corresponding 
$f$ is therefore given by,
\begin{equation}
f^N_2 = \frac{^{p}C_{q_2}} {^{q_1+q_2}C_{q_2}}
\label{subash}
\end{equation}
Note that these weights, wheather in the symmetrised form or unsymmetrised
form, are positive definite. Further, this is exactly the formula derived
by Chaturvedi and Srinivasan\cite{subash} in their microscopic analysis of
Haldane statistics for semions. 

It is important to stress the differences in these two approaches- in
their analysis Chaturvedi and Srinivasan start from a formulation of the
statistical mechanics of a system by removing factorizability of the
weights as a criterion. They derive the expression for the weights in 
eq.(\ref{subash}) by
imposing the conditions positivity and the requirement of symmetry (all
configuratons which are permutations of each other carry the same weight). 
Our starting point is the Sutherland model. We derive our rules from the
properties of shifted momenta. After removing this scaffolding, we obtain
not only positive definite weights for each configuration but when
symmetrised they reproduce the results of Chaturvedi and Srinivasan. 

\subsection{ The case with $g=1/3$ or $m=3$}
 
The maximal occupancy of a state in this case is 3. The allowed
configurations for each cell are (3),(21),(12),(111). That is we can form
a string of allowed configuration with any of these cells in any order to
make up $N$ particles. Any number of zeros may be added in between to make
up a total of $d$-states. 

As in the semion case we may consider expectation values of symmetric 
functions of $n_i$. Following the same procedure we can derive the 
symmetrized weights $f^N_3$ defined in eq.(\ref{fweight}).
Since, $m=3$, we have 
\begin{equation}
M_t(q_1,q_2,q_3) = ^{q_1+q_2+q_3}C_{q_3} ~~^{q_1+q_2}C_{q_2}, ~~~~ 
M_a(q_1,q_2,q_3) = ^{p}C_{q_3} ~~^{p-q_3}C_{q_2}~~ (2)^{q_2},
\end{equation}
where $p$, as before, is defined through the equation $N=3p+1$.  The 
corresponding weight $f$ is therefore given by,
\begin{equation}
f^N_3 = \frac{^{p}C_{q_3}~~^{p-q_3}C_{q_2} ~(2)^{q_2}}
{^{q_1+q_2+q_3}C_{q_3}~~ ^{q_1+q_2}C_{q_2}}
\end{equation}
These weights are again positive definite. Chaturvedi and 
Srinivasan\cite{subash} also suggest how their method may be extended 
beyond the semion case which they considered in detail. However, this 
extension requires additional conditions which are not imposed in the 
semion case. In contrast, our rules as derived from the point of view of 
an exactly solvable model are completely specified independent of the 
actual value of $g$ (or $m$). There is an algorithm to  derive 
$f^N_m$ for arbitrary $m$ though this gets complicated 
for larger $m$.

\section{Summary}

To summarise, we have analyzed the origin of negative probabilities in
exclusion statistics systems. To do this we have chosen an unusual
starting point in an equation and its solution given by Ramanujan. This
starting point makes precise the statements about the occurence of
negative probabilities. Further, we have formulated a counting principle
which reproduces the Haldane dimension formula. It can therefore be used
to define exclusion statistics purely in terms of state counting. The
negative probabilities discussed in literature can be understood as
arising when the system constrained by the counting rules is replaced by
an unconstrained one.  The negative weights then compensate for the
introduction of unphysical configurations. This is therefore exactly
analogous to other situations in physics where negative probablities
arise, for example, the ghosts and negative norm states in gauge theories 
or as in the case of Wigner distribution in quantum mechanics. 

{\bf Acknowledgement:} We are extremely grateful to S.K. Rangarajan for
drawing attention to the Ramanujam formula and pointing out its 
relevence.  We also thank S. Chaturvedi for discussions, clarifications 
and many helpful comments.

\end{document}